\documentclass[12pt,preprint]{emulateapj}

\shorttitle{RR\,Lyrae asteroseismology}
\shortauthors{Moln\'ar et al.}

\begin{document}

\title{Nonlinear asteroseismology of RR\,Lyrae}

\author{L. Moln\'ar\altaffilmark{1}, Z. Koll\'ath\altaffilmark{1}, R. Szab\'o\altaffilmark{1}, S. Bryson\altaffilmark{2}, K. Kolenberg\altaffilmark{3,4}, F. Mullally\altaffilmark{2,5}, S. E. Thompson\altaffilmark{2,5}}

\altaffiltext{1}{Konkoly Observatory, MTA CSFK, H-1121 Budapest, Konkoly Thege Mikl\'os \'ut 15-17., Hungary}
\altaffiltext{2}{NASA Ames Research Center, MS 244-30, Moffet Field, CA 94035, USA}
\altaffiltext{3}{Harvard-Smithsonian Center for Astrophysics, 60 Garden St., Cambridge MA 02138 USA}
\altaffiltext{4}{Institute of Astronomy, KU Leuven, Celestijnenlaan 200D, 3001 Heverlee, Belgium}
\altaffiltext{5}{SETI Institute, 189 Bernardo Ave Suite 100, Mountain View,
CA 94043, USA}

\email{molnar.laszlo@csfk.mta.hu}

\begin{abstract}
The observations of the \textit{Kepler} space telescope revealed that fundamental-mode RR\,Lyrae stars may show various radial overtones. The presence of multiple radial modes may allow us to conduct nonlinear asteroseismology: comparison of mode amplitudes and frequency shifts between observations and models. Here we report the detection of three radial modes in the star RR\,Lyr, the eponym of the class, using the \textit{Kepler} short cadence data: besides the fundamental mode, both the first and the ninth overtones can be derived from the data set. RR\,Lyrae shows period doubling, but switches occasionally to a state where a pattern of six pulsation cycles repeats instead of two. We found hydrodynamic models that show the same three modes and the period-six state, allowing for comparison with the observations.
\end{abstract}


\section{Introduction}
The precise and continuous observations of the \textit{Kepler} space telescope \citep{borucki} revealed a wealth of new features in RR\,Lyrae stars. Among the first discoveries was the detection of period doubling in three stars, including RR\,Lyrae itself \citep{kolenberg10, pd}. Additional frequencies in the Fourier transform, significant frequency components beside the usual RR\,Lyrae pattern (main period, its harmonics, modulation sidelobes and the Blazhko-modulation frequency) were also detected in several stars \citep{benko10, gug12}. Out of these new features, the period doubling phenomenon, the alternation of a higher and a lower amplitude pulsation cycle, has been successfully modeled. Our hydrodynamic calculations revealed that a high-order, 9:2 resonance can occur between the fundamental mode and the 9th radial overtone strange mode that leads to the period-doubling bifurcation \citep{kmsz11}. These results were also confirmed by \citet{smolec11}.

However, the nature of the additional frequencies is not fully understood yet, except for the ninth overtone which is related to the half-integer peaks and causes the period doubling. Some of the remaining peaks fall into the ranges where the first or second radial overtones are expected, but several stars display even more peaks, possibly indicating non-radial modes \citep{benko10, gug12}. Additional modes were also detected in stars observed by the CoRoT space telescope \citep{chadid10, poretti10, gug11}. The stars showing first or second overtone signals are unlike the usual double-mode RRd pulsators. The additional components fall into the mmag range so the amplitude ratios are extreme. 

Furthermore, almost all stars that show additional modes are Blazhko-variables. Only two stars have been found in the \textit{Kepler} sample that show a small second overtone with only traces of modulation down to the mmag level \citep{benko10, nemec11}. On the other hand, about half of the modulated stars in the Kepler sample show additional peaks in the Fourier spectrum, often together with signs of period doubling. Period doubling itself is not a regular process: it is modulated and sometimes shows signatures of additional bifurcations. The variations do not necessarily follow the Blazhko-modulation either \citep{pd}. Therefore it is necessary to investigate the origins of those variations.

\begin{figure*}
\includegraphics[scale=.85]{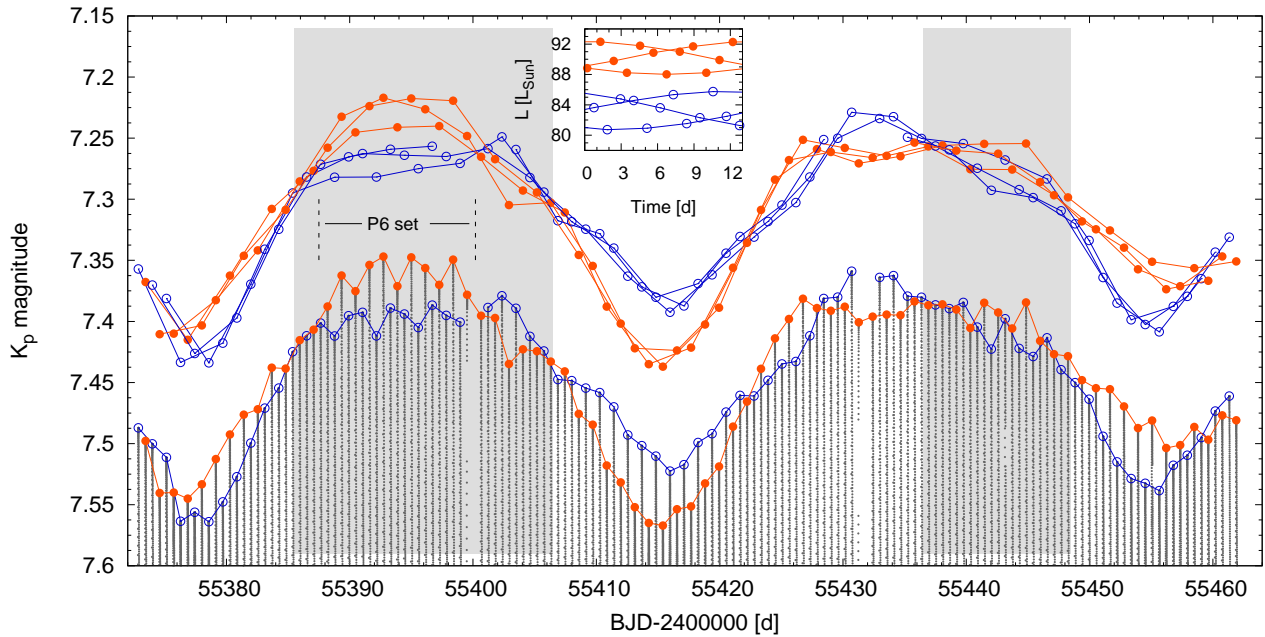}
\caption{Short cadence Q6 data of RR\,Lyrae. Light grey points are the brighter end of the light curve. Black and dark grey (blue and orange in the color version) lines connect the maxima of even and odd pulsation cycles respectively. In the case of the period-six visualization (shifted by 0.15 mag), every sixth maxima are connected, such as black/blue lines connect the $6k$, $6k+2$ and $6k+4$ points (circles), while grey/orange lines connect the $6k+1$, $6k+3$ and $6k+5$ points (dots) where $k=0,1,2$\dots. The period-six-like sections are indicated with the grey boxes: we identified two occurrences between 55385.3-55406.3 and 55436.9-55448.3. Dashed lines mark the P6 set which we analyzed separately (see Figure \ref{fou}). The insert shows the calculated luminosity of a model solution close to the 3:4 resonance. Note the similarity of the crossings between the individual blue and orange lines or branches. Model parameters ($\alpha_\nu$ is the eddy viscosity parameter) are: $M=0.64\, M_\odot$, $L=45\, L_\odot$, $T_{eff}= 6500\, K$, $\alpha_\nu=0.0175$. The insert covers the same time span as the P6 section.  }
\label{q5-q6}
\end{figure*}

\section{Kepler photometry of RR\,Lyrae}
RR\,Lyr (KIC\,7198959) is one of the brightest stars observed by \textit{Kepler} among its asteroseismology targets \citep{gilliland10}. It was observed in long cadence mode in \textit{Kepler} quarters 1 and 2 (or Q1 and Q2), with 29.4-minute sampling. \citet{kolenberg11} provided a detailed description of the photometry and the necessary corrections applied. The brightness of RR\,Lyr was initially underestimated in the Kepler Input Catalog (KIC) and some flux was lost from the less-than-ideal aperture. However, the corrected average brightness resulted in an assigned automatic aperture that extended beyond the edge of the CCD module and so the pipeline rejected the star in quarters 3 and 4. A custom aperture had to be developed for the star instead and it is being used from quarter 5 onwards. RR\,Lyrae was observed through the Kepler Guest Observer Program\footnote{http://keplergo.arc.nasa.gov/} in short cadence mode (1 minute sampling) both in Q5 and Q6.

Maxima of the pulsation cycles had to be determined for our analysis. For the long cadence data, we reused the maxima derived from ninth-order polynomials fitted in \citet{pd}. For the short cadence data we calculated spline solutions instead of polynomials. Long cadence data are less well suited to determine the sharp maxima of RR\,Lyrae-type variations, but we found the results satisfactory for further analysis. 

\subsection{Analysis of the photometry}
\label{dots}
In principle, return maps would be very useful to analyze the variations in period doubling detected in RR\,Lyrae stars. Return maps are powerful tools to examine the overall dynamics of a system by plotting the successive values of a selected property against each other, \textit{e.g.\ }the maximum values of the cycles of a (multi)periodic signal. On a return map, single-mode oscillation is represented by a single dot (no change), period doubling by two dots (small and large cycles), and double-mode pulsation is either a discrete number of points or a continuous loop, depending on whether the period ratio is commensurate or not. Return maps can provide an easier way to detect these features than conventional Fourier transforms as FTs of RR\,Lyrae variations are dominated by the harmonic peaks of the main period and may contain many linear combination peaks. 

In the case of Blazhko-variables however, the modulation will dominate over almost any other feature. So instead of creating a return map, we simply plotted the maxima of even and odd pulsation cycles respectively (blue and orange lines in Figure \ref{q5-q6}) in the light curve against time. If the period doubling is constant, the two lines should never cross each other. 

However, the pulsation of RR\,Lyr is much more complicated. The period doubling is omnipresent, contrary to the first implications from the LC data \citep{pd}, but it is clearly not constant. The most curious phenomenon occurs---coincidentally---in Q6, the 6th quarter of \textit{Kepler} observations, between BJD 55388 and 55400 (we refer to this section of the light curve as the P6 set, see Figure \ref{q5-q6}). An up-down bounce in each arm of period doubling is evident, indicating a possible additional bifurcation. Only if we connect every sixth maxima instead of every second can we smooth out most of the variations between successive maxima. Here the fundamental mode shows ``period-six" behavior, \textit{i.e.\ }a pattern of six cycles instead of two seems to be repeating. We found one more possible case and an extension to the P6 set where the pulsation temporarily shows traces of period-six behavior, indicated by the grey boxes on Figure \ref{q5-q6}. Outside the boxes, both visualizations (every second or sixth maxima) are dominated by period doubling, confirming the temporal nature of the period-six phenomenon. Reanalysis of the Q1-Q2 data suggests period-six pattern at the second half of Q1 and the beginning of Q2 but the gap between the quarters hampers the detection.

The period-six behavior can originate from a resonance if two modes are involved in the variation. A resonance with the first overtone with a period ratio of 3:4 (0.75) would be plausible for example, compared to the period ratio range of double-mode RR\,Lyrae stars \citep{petersen73}. We note in passing that it would be more correct to describe this resonance as 6:8 as the presence of the period doubling requires six pulsation cycles to repeat the pattern. Another possible mechanism is a periodic window in a chaotic regime. A period bifurcation cascade was indeed followed up to period-eight solutions in \citet{kmsz11}, but we have not found models yet where the period-doubling instability alone reached chaos. It is much more likely that the period-six behavior is connected to a resonance between the period-doubled fundamental mode and another overtone (which configuration itself can be a periodic window of stability between three-mode chaotic solutions).

\begin{figure}
\includegraphics[scale=.75]{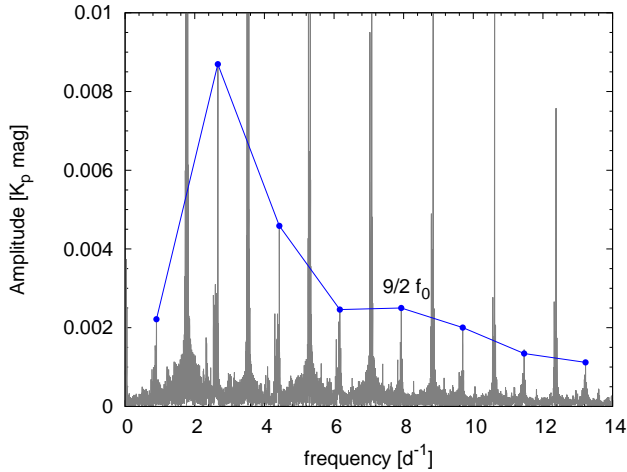}
\caption{Amplitudes of the highest half-integer peaks at every $(2n+1)/2 f_0$ value (blue points). The Fourier-spectrum of the Q5-Q6 data, prewithened with the $n f_0$ peaks, is shown in grey. There is an excess at $9/2 f_0$, at the position of the ninth overtone, similarly to Fig. 8 in \citet{pd}.}  
\label{hif}
\end{figure}

\subsection{The presence of the first radial overtone in RR\,Lyrae}
The analysis of the Q1--Q2 data of RR Lyr did not reveal any additional peaks beyond the half-integer frequencies \citep{kolenberg11}. The sign of the ninth overtone, locked in the 9:2 resonance, was detected in those data as an excess at $9/2 f_0$ \citep{pd}, and it is present in the Q5--Q6 short cadence data set too (Figure \ref{hif}). 

\begin{figure}
\includegraphics[scale=.85]{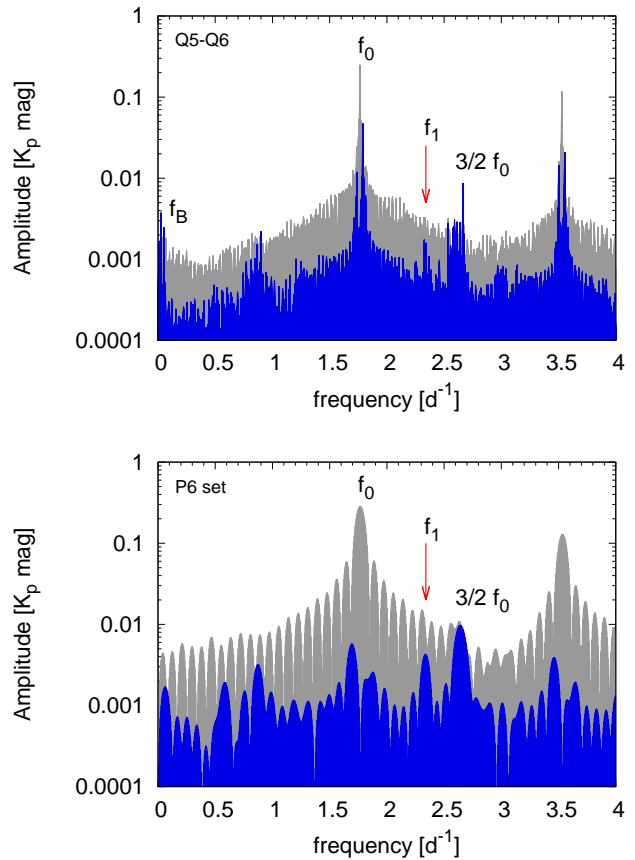}
\caption{Detection of the first overtone ($f_1$) in the \textit{Kepler} measurements of RR\,Lyr. The original spectra are in grey and the spectra prewhitened with the main frequency ($f_0$) and its harmonics ($nf_0$) are in black (blue in the color version). The upper panel shows the entire Q5--Q6 data while the lower panel shows only the P6 set from Q6 (see Figure \ref{q5-q6}). A forest of peaks is evident at $3/2 f_0$ in the upper panel, indicating the strong variability of the period doubling. } 
\label{fou}
\end{figure}

However, at least two significant peaks are  also detectable in the expected range of the first overtone in the Q5-Q6 data, at $f=2.327$ and 2.342 $d^{-1}$, with amplitude ratios ($A_0/A_1$, the ratio comparing the amplitude of the $f_0$ peak of the fundamental mode to the $f_1$ peak(s)) being 155 and 170, respectively. The presence of multiple peaks indicate amplitude variations and quite possibly phase variations as well. If analyzed separately, the Q5 and Q6 data result in somewhat different values, with $A_0/A_1=$ 138 and 102 for a sole significant peak. The spectrum of the P6 set (lower panel in Figure \ref{fou}) shows a stronger signal with $A_0/A_1=67$. The amplitudes are limited to a few mmags in all subsets, but thanks to the photometric precision, the individual peaks have signal-to-noise ratios of at least 4--6, the lowest values corresponding to the spectrum of the Q5 data only, where we did not detect period-six-like stages. 

Period ratios (or $f_0/f_1$ frequency ratios) vary between 0.753-0.758 for the different subsets and/or peaks: for the P6 set it is 0.756. These ratios are close to, but not exactly 0.750, the value corresponding to a 3:4 resonance. The period ratio difference might arise from the uncertainty in the frequency determination, either caused by the frequency variations connected to the Blazhko effect or, in the case of the P6 set, the shortness of the data set. But it is more likely that the system temporarily approaches the vicinity of the resonance and the frequency difference reflects the distance from it. An additional mode with the same ratio (0.754) was detected in UZ Vir too \citep{sodor12}.

This is the first time that three radial modes have been unambiguously detected in an RR\,Lyrae star: the fundamental, the first overtone and the third being the ninth overtone which is responsible for the period doubling. The detection is also strongly supported by the nonlinear models, described below. It is also clear however that the inclusion of period doubling and the first overtone still does not describe the light variations of RR Lyr perfectly. The two branches of the period doubling cross each other and reverse, with intervals ranging from a few pulsation periods to about 30 days which is close to an entire Blazhko-period ($\sim39$ days). These variations could be explained by yet another, physically relevant frequency close to the half-integer peaks (as it modulates the period doubling), or could be internal variability, as found in BL Herculis models \citep{sm12}.

\begin{figure}
\includegraphics[scale=.72]{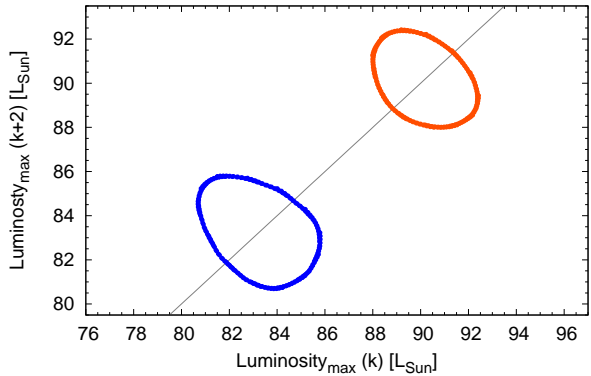}
\caption{First return map of a three-mode model. The figure shows maxima of the luminosity variations plotted against each other. Here we show the maxima of the $k$th versus the $(k+2)$th cycles instead of successive ones to separate the smaller and larger cycles of period doubling. Although three radial modes are present, the ninth overtone is locked in a resonance with the fundamental mode, and it is only detectable through the presence of period doubling, hence we observe two loops instead of a single one, both representing the double-mode pulsation between incommensurate $f_0$ and $f_1$ frequencies. Model parameters are the same as in Figure \ref{q5-q6}.}
\label{retmap}
\end{figure}

\section{Model calculations}
The modulations and possible bifurcations in period doubling raise further questions regarding the dynamics of the system. To study the the possible processes that might be responsible for this dynamical behavior we conducted a broad nonlinear model survey with the Florida-Budapest turbulent convective hydrodynamic code \citep{kollath01, kollath02}. The results of this survey will be published in a subsequent paper (Koll\'ath et al., in prep.), here we only summarize the relevant facts.

We found that period doubling changes the stability of the fundamental mode. The mode in itself is stable against the perturbations of the first overtone, but unstable against the ninth overtone and the period-doubling bifurcation occurs. The period-doubled mode is, however, unstable (or at least marginally stable) against the perturbations of the first overtone. The result is a new pulsation state that looks like a new kind of double mode behavior but is in fact created by the interaction of three different radial modes. In some of the models, this two-period  state bifurcates further to resonant or chaotic solutions. Detailed properties of these models, their occurrence along the instability strip will be investigated in forthcoming papers. We note that the two chaotic model solutions that were analyzed by \citet{plachy12} turned out to be three-mode solutions as well. 

The various dynamical states are best to represent with return maps. In case of simple period doubling, we will have two dots, corresponding to the lower and higher amplitude cycles. If two different modes are present, the return map will either show a set of points or a loop, depending on the phase relations (\textit{e.g.\ }are the two modes in resonance or not). 

Take for example the two loops in Figure \ref{retmap} which is the same model as in the insert in Figure \ref{q5-q6}. The presence of a loop at lower values and another at higher values indicate that both period doubling and an additional periodicity is present in this system, a state that resembles the Kepler observations of RR\,Lyrae.

\subsection{Comparison with the models}
The detection of three different radial modes in an RR Lyrae star may allow for nonlinear asteroseismic studies: comparison of the observations not only with linear pulsation periods but also with mode amplitudes, resonances and frequency shifts. 

Our first calculations focused on the possible 3:4 resonance. The corresponding 0.750 period ratio, and especially the exact observed period ratios in RR\,Lyr between the first overtone and the fundamental mode are a challenge: they lie at the edge of the accessible range for the classical double-mode pulsation in the Petersen-diagram (see Fig. 6 in \citet{szkb04}). We have not succeeded in finding models with an exact 3:4 resonance yet, though we found resonant solutions with higher integers corresponding to lower period ratios. These models can help to identify possible (near-)resonances between the fundamental mode and the first overtone in other RR\,Lyrae stars.

However, models with a linear period ratio of 0.747 instead of 0.750 do in fact approach the vicinity of the 3:4 resonance. These calculations indicate that RR\,Lyr only approaches the resonance during the period-six phases but does not lock into it. Although the models favor solutions below the 0.750 frequency ratio instead of above (as in the case of RR\,Lyr), the topology, like the distances and crossings of the six branches are quite similar, especially in the P6 section (see insert in Figure \ref{q5-q6}) and support our findings. Swapping of either three branches simultaneously was not observed in the models, confirming that yet another mechanism must be behind some of the remaining features, like the modulation of the period doubling.

We also note that the difference between linear and nonlinear period ratios in the models approaches 0.003 and sometimes (especially in resonant states) up to 0.005. This nonlinear shift is comparable to the observed period ratio range of the Petersen-diagram of the normal double-mode RR Lyrae (RRd) stars.

\section{Summary}
The detection of so many interesting features in a star of long observational history remarkably illustrates the advantages of space-based photometry. A decade ago, \citet{smith03} noted that the residuals after fitting the primary components of the spectrum (the triplet at the fundamental mode, their harmonics and the modulation frequency) exceeded the expected observational error. The first observations of \textit{Kepler} revealed the period-doubling phenomenon \citep{kolenberg10, pd} and now the first overtone was also detected in the short cadence data. The results and implications of the analysis can be summarized as follows:

\begin{itemize}
\item Three different radial modes were detected in RR\,Lyr, the eponym of its class: the fundamental mode and the first and ninth overtone. However, the ninth overtone is locked in a resonance with the fundamental mode and can be detected only through the presence of period doubling \citep{kmsz11}, causing the observed variations to resemble double-mode behavior.
\item The first overtone has a very small amplitude (few mmags), in contrast with the fundamental mode or the classical double-mode RRd stars, explaining the non-detection from the ground. 
\item A state resembling period-six bifurcation was detected in the light variations of RR\,Lyr. This behavior is most likely caused when the system approaches the vicinity of a 3:4 resonance between the fundamental mode and the first overtone.
\item We have been able to reproduce the three-mode state in nonlinear hydrodynamic models. Although we have not been able to model the 3:4 resonance or the higher than 0.750 period ratio of RR\,Lyr, the amplitude ratios can be reproduced and higher-order resonances have been found as well. Furthermore, near-resonance models below the 0.750 value have very similar characteristics, raising the hopes for applying nonlinear asteroseismology to Blazhko RR\,Lyrae stars.
\item Even the inclusion of three different modes cannot explain all the variations we see in the period doubling, indicating that the pulsation of RR\,Lyrae stars may be even more complex.
\end{itemize}

Our radial nonlinear hydrodynamic calculations demonstrated that the period-doubled fundamental mode may lose its stability with respect to the first radial overtone. Then it is straightforward to expect similar nonlinear destabilization with respect to low order nonradial modes as well. This process might provide the background for the fact that additional pulsation frequencies occur more frequently in stars where period doubling is also detected. First overtone signals were identified in three other stars in the \textit{Kepler} sample by \citet{benko10}, and the case of RR\,Lyr suggests that more may be discovered. The period ratios of peaks that can be plausibly associated with the first overtone are for V354 Lyr (KIC 6183128): $f_1/f_0=0.729$, and for V360 Lyr (KIC 9697825): $f_1/f_0=0.721$. For V445 Lyr (KIC 6186029), \citet{benko10} noted a very complex pattern of peaks while the detailed analysis of \citet{gug12} concluded that among the additional frequencies, two can be found at ratios $f_N/f_0=0.703$ and $f_0/f_1=0.730$. Even more stars show the signature of the second overtone for which \citet{molnar12_f2} already proposed the alternate idea of a three-mode resonance, where the equation $3f_0+f_2=f_9$ may describe the relation between the three frequencies. 

The results indicate that interactions between radial and nonradial modes may indeed play a more crucial role in the pulsation of stars. \citet{bk11} showed that the resonance responsible for period doubling can cause Blazhko-like modulation in amplitude equations. Nonlinear asteroseismology promises to understand these new findings, and might even help to unlock the mystery of the Blazhko effect itself.

\acknowledgments

Funding for this Discovery Mission is provided by NASA's Science Mission Directorate. The Kepler Team and the Kepler Guest Observer Office are recognized for helping to make the mission and these data possible. This work was supported by the Hungarian OTKA grants K83790, K81421 and MB08C 81013, as well as the `Lend\"ulet' programme, and the European Community's Seventh Framework Programme (FP7/2007-2013) under grant agreement no. 269194. RSz acknowledges the Bolyai J\'anos Scholarship of the Hungarian Academy of Sciences. KK is supported by a Marie Curie International Outgoing Fellowship within the $7^{\rm th}$ European Community Framework Programme. RSz and KK warmly thank the KITP staff of UCSB for their hospitality during the "Asteroseismology in the Space Age" programme. This research was supported in part by the National Science Foundation under Grant No. NSF PHY05-51164.

{\it Facilities:} \facility{Kepler ()}.

\end{document}